\documentclass{iau307}
\usepackage{graphicx}
\usepackage{natbib}
\usepackage{url}
\usepackage{dtklogos}
\bibpunct{(}{)}{;}{a}{}{,}

\title[Non-LTE abundances in the outer disk] %% give here short title %%
{Non-LTE Abundances in OB stars: Preliminary Results for 5 Stars in the Outer
 Galactic Disk
}

\author[Bragan\c ca et al.]   %% give here short author list %%
{Bragan\c ca, G. A.$^{1,2}$ \thanks{email: {\tt ga.braganca@gmail.com}},
 Lanz, T.$^2$,
 Daflon, S.$^1$,
 Cunha, K.$^{1,3}$,\\
 Garmany, C. D.$^4$,
 Glaspey, J. W.$^4$,
 Borges Fernandes, M.$^1$,\\
 Oey, M. S.$^5$,
 Bensby, T.$^6$,
 Hubeny, I.$^3$}

\affiliation{
$^1$Observat\'orio Nacional, Rio de Janeiro, Brazil;
$^2$Observatoire de la C\^ote d'Azur, Nice, France;
$^3$Steward Observatory, Tucson, AZ, U.S.A.;
$^4$National Optical Astronomy Observatory, Tucson, AZ, U.S.A;
$^5$University of Michigan, Ann Arbor, MI, U.S.A.;
$^6$Lund Observatory, Lund, Sweden}

\pubyear{2014}
\volume{307}
\pagerange{}
\jname{New windows on massive stars: asteroseismology, interferometry, and spectropolarimetry}
\editors{G. Meynet, C. Georgy, J.H. Groh \& Ph. Stee, eds.}
\begin{document}

\maketitle

\begin{abstract}
The aim of this study is to analyse and determine elemental abundances for a
 large sample of distant B stars in the outer Galactic disk in order to
 constrain the chemical distribution of the Galactic disk and models of
 chemical evolution of the Galaxy.
 Here, we present preliminary results on a few stars along with the adopted
 methodology based on securing simultaneous O and Si ionization equilibria with
 consistent NLTE model atmospheres.
\end{abstract}

The chemical distribution of B stars in the outer Galactic disk is presently
 poorly probed, based on only a few abundance results for B distant stars
 (e.g., \citealt{Daflon04}). In order to enlarge the number of studied stars
 and to better represent the chemical distribution of the outer Galactic disk,
 we obtained  high-resolution echelle spectra for a sample of 136 OB stars
 located towards the Galactic anti-center using the MIKE spectrograph on the
 6.5m Magellan Clay telescope. A subsample of 50 sharp-lined B stars has been
 selected for the abundance analysis. High resolution, high signal-to-noise
 spectra of 3 well studied main-sequence B stars (HD 61068, HD 63922 and HD
 74575) were added to the sample in order to test our adopted methodology.

We use an iterative method to obtain simultaneously the stellar parameters
 (effective temperature ($T_{eff}$), surface gravity ($\log g$),
 microturbulence, and abundances of Si and O) based on non-LTE synthesis of H,
 He, Si, and O profiles.  The synthetic spectra are computed using
 \texttt{SYNSPEC} \citep{SYNSPEC}, which interpolates in a grid of non-LTE
 model atmospheres computed with \texttt{TLUSTY} \citep{TLUSTY} and detailed
 atomic models of O (69, 219 and 41 levels for O I, II and III, respectively)
 and Si (70, 122 and 53 levels for Si II, III and IV, respectively). The
 adopted method, based on \cite{Hunter07}, consists of the following steps:

\begin{enumerate}
\item Initial values for the stellar parameters are set from
 the stellar spectral type.
\item Ionization balance of Si II/III/IV and/or O I/II/III provides  $T_{eff}$
 and abundances of Si and O.
\item $\log g$ is obtained from fits of the pressure broadened wings of the
 Balmer lines H$\alpha$, H$\beta$ and H$\gamma$.
\item Microturbulence is defined by requiring that the Si III line abundances
 are independent of the line strength (equivalent width).
\item We check for convergence of the basic stellar parameters: $T_{eff}$,
 $\log g$ and microturbulence. If not, the previous steps are repeated.
\item If converged, we fit the Si and O lines to obtain the adopted abundance
 values.
\end{enumerate}
The parameters $T_{eff}$, $\log g$, microturbulence and Si and O abundances
 (in the $\log \epsilon_X = \log(N_X/N_H) +12$ notation) are obtained with
 uncertainties of  325 K, 0.07 dex, 3 km/s, 0.08 dex and 0.11 dex, respectively.

\begin{figure}[b]
\begin{center}
\includegraphics[width=\textwidth]{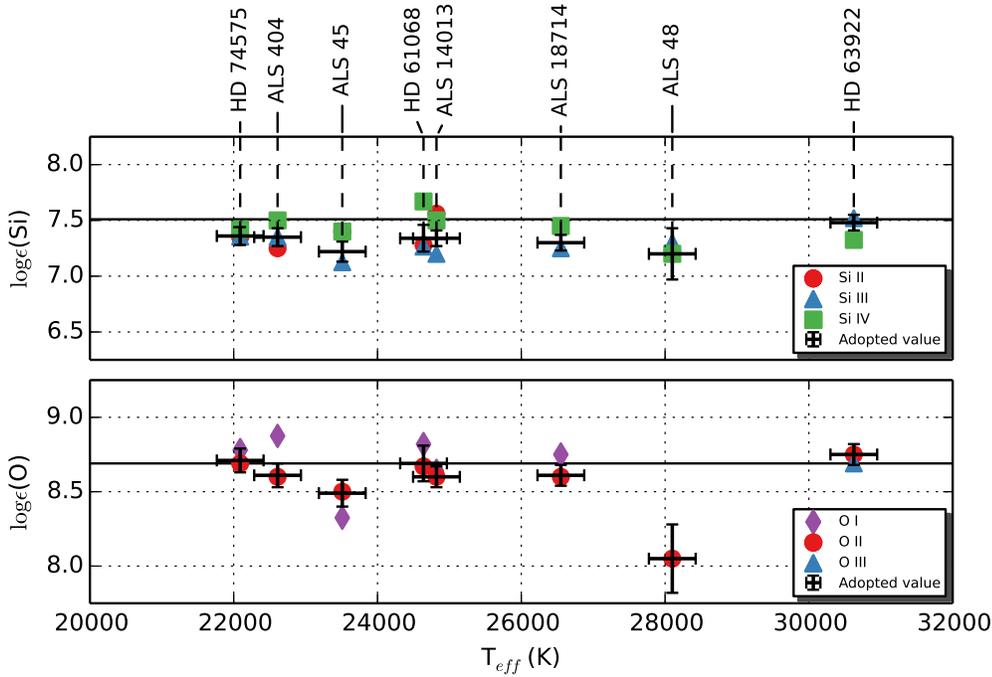}
\caption{Chemical abundances of Si and O as a function of $T_{eff}$.
 The black crosses represent the adopted abundances for Si and O with
the respective dispersion. The colored symbols show the abundance obtained for
 each species (see legend) and  the solid lines represent the solar value
 obtained by \cite{Asplund09}.}
\label{fig1}
\end{center}
\end{figure}

Our preliminary abundance results and effective temperatures for 5 ALS sample
 stars and 3 reference HD stars are shown in Figure \ref{fig1}. The derived
 abundances show no trends with effective temperature, indicating the non-LTE
 calculations are free of systematics. This iterative method will be applied to
 the whole sub-sample of 50 sharp-lined, distant B stars in order to probe the
 chemical distribution  of the outer Galactic disk.

\textbf{Acknowledgements:} We acknowledge financial support of National Science
 Foundation (NSF, AST-0448900), Coordena\c c\~ao de Aperfei\c coamento de
 Pessoal de N\'ivel Superior (CAPES) and the International Astronomomic Union
 (IAU).
\bibliographystyle{iau307}
\bibliography{MyBiblio}

\begin{thebibliography}{}

\bibitem[\protect\astroncite{{Asplund} et~al.}{2009}]{Asplund09}
{Asplund}, M., {Grevesse}, N., {Sauval}, A.~J., \& {Scott}, P. 2009,
\newblock {\em \araa} 47, 481

\bibitem[\protect\astroncite{{Daflon} et~al.}{2004}]{Daflon04}
{Daflon}, S., {Cunha}, K., \& {Butler}, K. 2004,
\newblock {\em \apj} 606, 514

\bibitem[\protect\astroncite{{Hubeny} \& {Lanz}}{1995}]{TLUSTY}
{Hubeny}, I. \& {Lanz}, T. 1995,
\newblock {\em \apj} 439, 875

\bibitem[\protect\astroncite{{Hubeny} \& {Lanz}}{2011}]{SYNSPEC}
{Hubeny}, I. \& {Lanz}, T. 2011,
\newblock {\em {Synspec: General Spectrum Synthesis Program}},
\newblock Astrophysics Source Code Library

\bibitem[\protect\astroncite{{Hunter} et~al.}{2007}]{Hunter07}
{Hunter}, I., {Dufton}, P.~L., {Smartt}, S.~J., {et~al.} 2007,
\newblock {\em \aap} 466, 277

\end{thebibliography}

\end{document}